\documentclass[nofootinbib,superscriptaddress]{revtex4-1}
\usepackage{epsfig}
\usepackage{multirow}
\usepackage{amsmath}
\usepackage{amssymb}
\begin{document}

%\preprint{}

\title{Structural effects of $^{34}$Na in the $^{33}$Na(n,$\gamma)^{34}$Na radiative capture reaction}


\author{G. Singh}
\email{gagandph@iitr.ac.in}
\affiliation{Department of Physics, Indian Institute of Technology - Roorkee, 247667, INDIA}
\author{Shubhchintak}
\email{shub.shubhchintak@tamuc.edu}
\affiliation{Department of Physics and Astronomy, Texas A$\&$M University - Commerce, 75429, USA}
%\altaffiliation{\it }
\author{R. Chatterjee}
\email{rcfphfph@iitr.ac.in}
%\thanks{}
\affiliation{Department of Physics, Indian Institute of Technology - Roorkee, 247667, INDIA}


\date{\today}

\begin{abstract}
%An article usually includes an abstract, a concise summary of the work
%covered at length in the main body of the article. 
\begin{description}
\item[Background]
The path towards the production of \textit{r}-process seed nuclei follows a course where the neutron rich light and medium mass nuclei play a crucial role. The neutron capture rates for these exotic nuclei could dominate over their $\alpha$-capture rates, thereby enhancing their abundances at or near the drip line. Sodium isotopes especially should have a strong neutron capture flow to gain abundance at the drip line. In this context, study of $^{33}$Na(n,$\gamma)^{34}$Na and $^{33}$Na($\alpha$,n)$^{36}$Al reactions becomes indispensable.
\item[Purpose]
In this paper, we calculate the radiative neutron capture cross-section for the $^{33}$Na(n,$\gamma)^{34}$Na reaction involving deformation effects. Subsequently, the rate for this reaction is found and compared with that of the $\alpha$-capture for the $^{33}$Na($\alpha$,n)$^{36}$Al reaction to determine the possible path flow for the abundances of sodium isotopes.
\item[Method]
We use the entirely quantum mechanical theory of finite range distorted wave Born approximation upgraded to incorporate deformation effects, and calculate the Coulomb dissociation of $^{34}$Na as it undergoes elastic breakup on $^{208}$Pb when directed at a beam energy of 100 MeV/u. Using the principle of detailed balance to study the reverse photodisintegration reaction, we find the radiative neutron capture cross-section with variation in one neutron binding energy and quadrupole deformation of $^{34}$Na. The rate of this $^{33}$Na(n,$\gamma)^{34}$Na reaction is then compared with that of the $\alpha$-capture by $^{33}$Na deduced from the Hauser-Feshbach theory.
\item[Results]
The non-resonant one neutron radiative capture cross-section for $^{33}$Na(n,$\gamma)^{34}$Na is calculated and is found to increase with increasing deformation of $^{34}$Na. An analytic scrutiny of the capture cross-section with neutron separation energy as a parameter is also done at different energy ranges. The calculated reaction rate is compared with the rate of the $^{33}$Na($\alpha$,n)$^{36}$Al reaction, and is found to be significantly higher below a temperature of $T_9 = 2$.
\item[Conclusion]
%Our computations manifested that in the relevant range of the astrophysical energy and temperature, the cross-section and rate of the radiative neutron capture increased with increasing deformation and decreasing one neutron separation energy of $^{34}$Na. Moreover, 

At the equilibrium temperature of $T_9 = 0.62$, the rate for the neutron capture had a small but non-negligible dependence on the structural parameters of $^{34}$Na. In addition, this neutron capture rate exceeded that of the $\alpha$-capture reaction by orders of magnitude, indicating that the $\alpha$-process should not break the (n,$\gamma$) \textit{r-}process path at $^{33}$Na isotope, thus, effectively pushing the abundance of sodium isotopes towards the neutron drip line.
 
\end{description}
\end{abstract}

%\begin{abstract}
%We calculate the Coulomb dissociation of $^{16}$N on a Pb target at 100 MeV/u incident beam energy within the fully quantum mechanical distorted wave Born approximation formalism of breakup reactions. The capture cross section and subsequently rate of the $^{15}$N($n,\gamma$)$^{16}$N reaction are calculated from the photodisintegration of $^{16}$N, using the principle of detailed balance. Our theoretical model is free from the uncertainties associated with the multipole strength distributions of the projectile.

%\end{abstract}

%\keywords{}

\maketitle

\section {Introduction}
\label{sec:1}
The explanation of the abundance curve has been an enigma for more than half a century. The formation of light to medium mass nuclei could be accounted for from the results of hydrostatic nucleosynthesis, but the energy economics alone could not explain the endothermic reactions required for the elemental production for \textit{A} $\gtrsim$ 60. It was postulated that various nucleosynthesis processes (viz., the \textit{pp-}chains, the CNO-cycles, \textit{p-, s-, rp-, r-} processes) occurred in stellar plasma under different physical conditions resulting in the formation of the elements found today in our universe \cite{BBFH, Rolf, 40BBFH, Iliadis}. The \textit{pp-}chains are a series of fusion reactions for hydrogen nuclei fusing together to form an $\alpha$-particle and are the most probable energy sources in main sequence stars, while in the CNO-cycles, four hydrogen nuclei fuse stimulated by carbon, nitrogen and oxygen, to emit an $\alpha$-particle, two positrons and two electron neutrinos \cite{Wiescher}. The \textit{p-}process is speculatively responsible for the formation of proton rich elements with \textit{A} $\simeq$ 100 that are inhibited production by the \textit{s-} or \textit{r-}processes because of the occurrence of stable nuclei in their paths \cite{40BBFH, Hoffman} and the genesis of heavier elements beyond iron in a highly proton dominating environment at temperatures higher than those found in the main sequence stars is attributed mainly to the \textit{rp-}process \cite{Wallace}. It differs from the \textit{p-}process in that it occurs close to the proton drip line and is identical to the \textit{r-}process on the neutron rich side except for the Coulomb barrier. Due to the very short life time of neutron-capture reactions relative to $\beta$-decay, the \textit{rapid} neutron-capture or the \textit{r}-process (unlike the \textit{slow} neutron-capture or the \textit{s}-process) occurs far from the valley of stability resulting in low binding energy of the nuclei. It is believed to be responsible for most of the nuclei and atoms heavier than iron in this region. Though it is known that the \textit{r}-process occurs under explosive conditions of temperature and pressure, the exact astrophysical sites for its occurrence are still not conclusive \cite{Cowan, Meyer, Thielemann, Tanvir, Qian, Mennekens, Bauswein, Shen, Voort, Lippuner}. The uncertainty in determining the exact sites for the \textit{r}-process can also, in part, be attributed to lack of experimental data available for the relevant neutron rich nuclei. 

\textit{r}-process nucleosynthesis calculations are also known to include neutron rich light and medium mass nuclei in their reaction networks, for their exclusion can change heavy element abundances considerably \cite{Tera,Sasaqui}. In a He-rich environment, $\alpha$-capture reactions should essentially dominate at higher temperatures and densities. This gives rise to a competition amongst $\alpha$-capture, $\beta$-decay and neutron capture reactions participating in the \textit{r}-process \cite{Sasaqui}. Far from the valley of stability towards the neutron rich side, if there is an equilibrium between (n, $\gamma$) and ($\gamma$, n), theoretically, the \textit{r}-process paths should lead up to the drip line isotope. Nevertheless, the contention between $\alpha$-capture and neutron capture ensures that domination of $\alpha$-capture should potentially break the \textit{r}-process flow of radiative neutron capture followed by a $\beta$-decay. This will result in the promotion of the atomic number, \textit{Z}, of the nucleus and the isotope production in the same (n, $\gamma$)-($\gamma$, n) chain is reduced.

It has been reported \cite{Tera} that under a short dynamic time scale model, such light and medium mass nuclei very near or at the drip line have shown largest abundances for each atomic number, \textit{Z}, except for the isotopes $^{18}$C and $^{36}$Mg, which are comparatively away from their respective drip nuclei. These abundance patterns can be predicted by studying reaction rates for different reactions that the nuclei might be involved in and comparing them with observations. It is, therefore, imperative that one knows the correct neutron and $\alpha$-capture rates for the light and medium mass nuclei in the `island of inversion' \cite{War} (\textit{N} = 20-30), so as to predict the correct abundance patterns and availability of nuclei as participants or seeds in the \textit{r}-process.

Following this abundance pattern, $^{35}$Na is the most abundant sodium isotope near the neutron drip line \cite{Tera}, whose production will depend largely on the abundance of $^{34}$Na: its ground state (g.s.) spin-parity and binding energy and its availability to form $^{35}$Na. This, in turn, should depend on reaction rates determining the formation of $^{34}$Na and its subsequent decay. If the reaction rate for the $^{33}$Na($\alpha$,n)$^{36}$Al capture reaction is higher than the rate of $^{33}$Na(n,$\gamma)^{34}$Na reaction, the reaction network will follow a different path and formation of $^{34}$Na will be retarded. It is also interesting to note that $^{34}$Na lies in the deformed medium mass region (\textit{N} = 20-30), where exotic nuclei have been found recently \cite{Moto,Naka2009,31Ne,37Mg,Doornenbal}. Deformation in this nucleus can affect its cross-section as well as its one neutron separation energy and ground state spin-parity: parameters that can greatly influence its abundance \cite{Goriely}.

The aim of this paper is to report the findings of our investigations on the rate of $^{33}$Na(n,$\gamma)^{34}$Na capture reaction and compare it with the $^{33}$Na($\alpha$,n)$^{36}$Al capture, at stellar energies corresponding to the astrophysically relevant temperatures (T$_9$ = 0.5 - 10; $T_9 = 1$ corresponds to a temperature of $10^9$ K). This is significant because according to Ref. \cite{Tera}, neutron captures by these light and medium mass seed nuclei from the line of $\beta$-stability till the neutron drip line will diminish the number of neutrons available to make heavier nuclei. At the equilibrium temperature (T$_9$ = 0.62) and mass density ($\rho = 5.4 \times 10^2$g/cc), Na isotopes are supposed to maintain a very strong flow of neutron capture, pushing isotope formation near the drip line \cite{Tera}. This should ideally result in $^{33}$Na(n,$\gamma)^{34}$Na having a larger reaction rate than $^{33}$Na($\alpha$,n)$^{36}$Al, a dictum which can be confirmed only by meticulous and accurate determination of these reaction rates. 

The relevant temperature range (T$_9$ = 0.5 - 10) roughly equates to a centre of mass energy range of about 50 keV to 1 MeV. At such low energy range and because of the acutely small half-life of $^{33}$Na ($\approx$ 8 ms \cite{DataA33}), performing direct reaction experiments is extremely difficult. Therefore, indirect methods have to be used to calculate cross-sections and reaction rates at these energies. We use the elegant indirect method of Coulomb dissociation (CD) \cite{BaurBert, 9Li} to probe the $^{33}$Na(n,$\gamma)^{34}$Na reaction, theoretically. Coulomb dissociation involves breakup of a projectile into a core and (a) valence nucleon(s) due to its dynamics in the electromagnetic field of a stable heavy target. CD is advantageous in the sense that it allows an inspection even at low relative energies of the final channel fragments despite the fact that it can be applied even to higher beam energy measurements keeping the target in its ground state \cite{Nakamura}.

We assume elastic dissociation of $^{34}$Na into a $^{33}$Na core and a valence neutron in the Coulomb field of a heavy $^{208}$Pb target. The theory of finite range distorted wave Born approximation (FRDWBA) extended to include deformation effects in the projectile is applied to extract the relative energy spectra for the breakup reaction. FRDWBA is a fully quantum mechanical theory which only requires the full ground state projectile wave function as an input. It has an added advantage over first order theories in that it covers the target-projectile electromagnetic interaction to all orders and the breakup contributions from the entire non-resonant continuum. Therefore, it is free from the uncertainties associated with multipole strength distributions which occur in many other theoretical models \cite{9Li,Goriely,BHTreview} \footnote{Of course, this assertion has to be qualified by stating that in the post form reaction theory there should not be any resonant structures in the core-valence particle/cluster continuum \cite{BHTreview}.}. 
Calculating the photodisintegration cross-section from FRDWBA for the breakup of $^{34}$Na, we then summon the principle of detailed balance to calculate the capture cross-section for the reverse reaction \cite{BertJPG,9Li} and utilize it to find the relevant reaction rates. Since the one neutron separation energy ($S_n$) and the quadrupole deformation ($\beta_2$) values are not fully established for $^{34}$Na \cite{Gaudefroy, 34Na}, we also study the variation of these capture cross-sections and rates keeping $S_n$ and $\beta_2$ as parameters. The behaviour of neutron capture rates for the uncertain ground state spin of $^{34}$Na is also discussed. Eventually, we compare the rate of the $^{33}$Na(n,$\gamma)^{34}$Na capture reaction with that of $^{33}$Na($\alpha$,n)$^{36}$Al reaction as obtained from the Hauser-Feshbach (HF) model using the NON-SMOKER code \cite{NS} and conclude that at the physical conditions specified, the probability of a neutron capture is greater than that of an $\alpha$-capture.

% - and thus, only a fixed depth of the potential required to reproduce the binding energy for the projectile ground state -

In the next Section we present our formalism, while in Section \ref{sec:3} we discuss our results. Section \ref{sec:4} highlights the conclusions.

\section{Formalism}
\label{sec:2}
To explore the prospective role of $^{34}$Na in the \textit{r}-process and in the elemental abundance near the drip line, we study its Coulomb dissociation (CD) on a heavy target and use the observables so obtained to calculate the rate of the $^{33}$Na(n,$\gamma)^{34}$Na capture reaction at stellar temperatures. The method we use for CD studies is the FRDWBA which has been advanced to include the effects of deformation on a nucleus \cite{31Ne}.

Let us contemplate a beam of projectile \textit{a} ($^{34}$Na) impinging on a heavy target \textit{t} ($^{208}$Pb) at 100 MeV/u. The reaction $a + t \rightarrow b + c + t$ occurs due to the heavily repulsive Coulomb field of $^{208}$Pb which excites the ($^{34}$Na) projectile above its particle emission threshold such that it undergoes elastic Coulomb breakup and a core, \textit{b} ($^{33}$Na) and a valence nucleon, \textit{c} (neutron) are ejected. Using the three-body Jacobi coordinate system (shown in Fig. 1 ), we write the relative energy spectrum for the breakup of this two-body composite system ($^{34}$Na) as \cite{9Li}:

\begin{figure}[ht]
\centering
\includegraphics[height=4.1cm, clip,width=0.40 \textwidth]{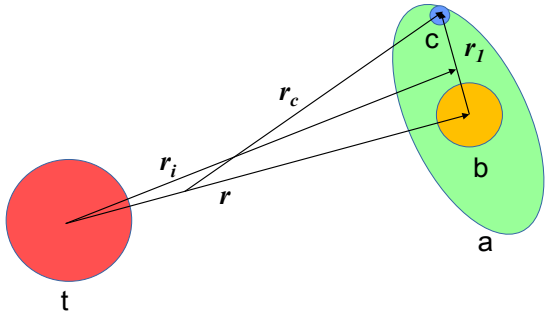}
\caption{\label{fig: Jacobi} (Color online) The three-body Jacobi coordinate system. The \textbf{\textit{r}}'s are the corresponding position vectors \cite{34Na}.}
\end{figure}

\begin{eqnarray}
\dfrac{d\sigma}{dE_{bc}} &=&  
\iint \left(\dfrac{\mu_{at}\mu_{bc}p_{at}p_{bc}}{(2\pi)^{5}\hbar^{7} v_{at}}\right)\left\lbrace \sum_{lm}\dfrac{1}{(2l+1)}|\beta_{lm}|^{2} \right\rbrace d\Omega_{at}d\Omega_{bc}, 	\label{a1.1}
\end{eqnarray}

where, $\Omega$'s are the solid angles, $\mu$'s are the reduced masses and \textit{p}'s are the appropriate linear momenta corresponding to the respective two-body systems. $v_{at}$ is the \textit{a-t} relative velocity in the entry channel, $E_{bc}$ is the relative or the centre of mass energy of the \textit{b-c} system (it will be used interchangeably as $E_{c.m.}$). $l$ and $m$ are the relative orbital angular momentum of the \textit{b-c} system, and its projection, respectively.

For a nucleus near the neutron drip line, particle \textit{c} is a neutron (case in point), which renders the reduced transition amplitude, $\beta_{lm}$, of Eq. \ref{a1.1} to take the form \cite{RC1}:

\begin{eqnarray}
\hat{l}\beta_{lm} = \int d\textbf{r}_{i}e^{-i\delta\textbf{q}_{c}.\textbf{r}_{i}} \chi_{b}^{(-)*}(\textbf{q}_{b},\textbf{r}_{i})\chi_{a}^{(+)}(\textbf{q}_{a},\textbf{r}_{i}) \int d\textbf{r}_{1}e^{-i(\gamma\textbf{q}_c-\alpha\textbf{K}).\textbf{r}_{1}}V_{bc}(\textbf{r}_{1})\phi_{a}^{lm}(\textbf{r}_{1}). 	\label{a1.2}
\end{eqnarray}

Here, $\alpha, \gamma$ and $\delta$ are the mass factors according to the Jacobi coordinate system while \textbf{q}'s are the Jacobi wave vectors corresponding to the respective nuclei; $\textbf{K}$ is the effective local momentum for the core-target system \cite{31Ne}. The $\chi$'s are taken to be pure Coulomb distorted waves whose convolution with the plane wave for particle \textit{c} in the first integral in Eq. (\ref{a1.2}) describes the dynamics of the reaction. The second integral expresses the structure part of the reaction by involving the ground state wave function of the projectile ($\phi_{a}^{lm}(\textbf{r}_{1})$) and the potential, $V_{bc}(\textbf{r}_{1})$. The deformation is incorporated in our FRDWBA theory via this axially symmetric quadrupole-deformed potential, which is constructed as \cite{31Ne}:

\begin{eqnarray}
V_{bc}(\textbf{r}_{1}) = V_{s}(r_{1}) - \beta_{2}V_{ws}
R\left[ \frac{d{g(r_{1})}}{dr_{1}} \right] Y_{2}^{0}(\hat{\textbf{r}}_{1}), \label{a1.3}
\end{eqnarray}

where \textit{V$_{ws}$} is the Woods-Saxon potential depth, $\beta_{2}$ is the quadrupole deformation parameter and \textit{g}(r$_{1}$) = $\left[{1 + exp(\frac{r_{1} - R}{a})}\right]^{-1}$ with radius $R = r_{0}A^{1/3}$. $A$ is the mass number of the projectile and $r_{0}$ and $a$ are the radius and diffuseness parameters, fixed at 1.24 fm and 0.62, respectively. It is worth noting that although we have used a deformed potential to define the interaction between final projectile fragments, we have used a ground state wave function from a spherical Woods-Saxon potential, $V_s(r_1)$ (given by $V_{ws}\times g(r_{1})$). This may sound contradictory at first, but it has been shown that for weakly bound nuclei with very low separation energies, the contribution from higher orbital angular momenta gets suppressed and only the lower $l$ values contribute significantly \cite{Hama, 34Na}. Thus, one might use the ground state wave function from a spherical potential for such cases until finer mathematical developments occur for the implementations required to remove this approximation.\footnote{In retrospect, calculations with a fully deformed ground state wave function of the projectile would be welcome. In fact, in Ref. \cite{VinhMau}, studies for the effect of particle-vibration coupling on single neutron states have been done for light halo nuclei. These couplings are believed to be responsible for the inversion of $1/2^-$ - $1/2^+$ levels in $^{11}$Be. Such calculations would indeed be interesting if carried out in the medium mass region, especially for the so called `island of inversion', which could also help in constraining the spectroscopic factors \cite{KobayashiPRL,37Mg}.}

Now, if transitions of a single multipolarity and type dominate the breakup cross-section and the nuclear breakup effects can be ignored, the relative energy spectrum of the three-body elastic Coulomb breakup obtained from Eq. (\ref{a1.1}) above can be used to obtain the total photodisintegration cross-section as \cite{BaurBert},
%\footnote{ The relative energy spectra is related to the two-body photodisintegration cross-section ($\sigma_{\gamma,n}$) via the equation \cite{BaurBert} $\dfrac{d\sigma}{dE_{bc}} = \dfrac{1}{E_{\gamma}}\sum_{\lambda} n_{\pi\lambda}\sigma_{(\gamma, n)},$}:

\begin{eqnarray}
\sigma_{(\gamma, n)} = \left(\dfrac{d\sigma}{dE_{bc}}\right) \left(\dfrac{E_{\gamma}}{n_{E1}}\right)  \label{a1.4}
\end{eqnarray}

since in the case of $^{33}$Na(n,$\gamma)^{34}$Na reaction, transitions of multipolarity $E1$ should dominate \cite{34Na}. $E_{\gamma}$ is the sum of the relative energy in the centre of mass (c.m.) frame between the core-valence neutron ($E_{bc}$ or $E_{c.m.}$) and the valence neutron binding energy ($S_n$). $n_{E1}$ is the virtual photon number for electric dipole transitions \cite{BertBaur, BertJPG}.

The principle of detailed balance states that each process should be equilibrated by its reverse process at equilibrium, which means that the capture cross-section for the $^{33}$Na(n, $\gamma)^{34}$Na reaction can be calculated from the time reversed $^{34}$Na($\gamma$, n)$^{33}$Na reaction via \cite{RolfsRodney}:

\begin{eqnarray}
\sigma_{(n, \gamma)} = \dfrac{2{\hat{j_a}}^2}{{\hat{j_b}}^2{\hat{j_b}}^2}\dfrac{k_{\gamma}^2}{k_{bc}^2}\sigma_{(\gamma, n)}	\label{a1.5}
\end{eqnarray}

where, ${\hat{j_i}}^2 = (2j_i + 1): j_i$ is the spin of the $i^{th}$ particle; $i \in \{a, b, c\}$. $k_{\gamma}$ is the photon wave number and $k_{bc}$ is the wave number of the relative motion between \textit{b} and \textit{c}. Thus, knowing the photodisintegration cross-section for a reaction can give us the radiative capture cross-section for its inverse reaction.
%This is a well known formalism to calculate cross-sections and reaction rates for nuclei at the relevant stellar energy ranges \cite{9Li,BertJPG}.

For non-degenerate stellar matter, the rate of a nuclear reaction ($R$) for two nuclei forming a composite system via the radiative capture process is given by \cite{RolfsRodney}:

\begin{eqnarray}
R = N_{A}\langle\sigma(v_{bc})v_{bc}\rangle 	\label{a1.6}
\end{eqnarray}

where, $N_A$ is the Avogadro number and $v_{bc}$ is the relative velocity corresponding to the c.m. energy $E_{c.m.}$. The product $\sigma(v_{bc})v_{bc}$ is the non-resonant reaction rate per particle pair and is averaged over the Maxwell-Boltzmann velocity distribution. It is defined as \cite{RolfsRodney}:

\begin{eqnarray}
\langle\sigma(v_{bc})v_{bc}\rangle &=& \sqrt{\dfrac{8}{\pi\mu_{bc}(k_BT)^3}}\int_0^\infty dE_{bc}\sigma_{(n, \gamma)}(E_{bc})E_{bc}exp(-\frac{E_{bc}}{k_BT}) 	\label{a1.7}
\end{eqnarray}

with $k_B$ being the Boltzmann constant and $T$, the stellar temperature, which, in nuclear astrophysics, is usually taken in units of $T_9$. Hence, knowing the relative energy spectrum of a single multipole dominated reaction from CD studies, and using Eqs. (\ref{a1.4}), (\ref{a1.5}) and (\ref{a1.6}), we can easily calculate reaction rates of stellar reactions in this elegant indirect manner \cite{9Li,16N}. Such indirect approaches are used quite extensively in nuclear astrophysics for studying a diversity of nuclear reactions and obtaining information about the events occurring in the stellar plasma \cite{Casal, Tribble, ShubhIOP}. They are essential because given the experimental technologies available, the direct measurement of radiative capture cross-sections, $\sigma_{(n,\gamma)}(E_{bc})$, for most astrophysical sites is difficult at such low ranges of relative energy ($E_{bc} \sim 10^{-3}$ \textendash 1 MeV).

We must maintain however, that the above method is only applicable when the breakup cross-section is dominated by transitions of a single multipolarity and type and the higher order effects contributing to the CD cross-sections are negligible at the beam energies considered \cite{BBHST}. For more details on the formalism, one may refer to \cite{9Li, 16N}.
%%%%%%%%%%%%%%%%%%%%%%%%%%%%%%%%%%%%%%%%%%%%%%%%%%%%%%%%%%%%%%%%%%%%%%%%%%

\section{RESULTS AND DISCUSSION}
\label{sec:3}

Theoretical investigations for different observables in the elastic Coulomb breakup of $^{34}$Na have been done in Ref. \cite{34Na} and they have suggested it to have a halo structure, with its ground state configuration possibly being $^{33}$Na$(3/2^{+})$ $\otimes$ \textit{$2p_{3/2}\nu$}. It is shown that the peak position of the relative energy spectrum changes with changing deformation and the effect of deformation on scaling laws has also been discussed. The ground state spin-parity of $^{34}$Na is uncertain: it could be $0^-, 1^-, 2^-$ or $3^-$. Shell model predictions put its $J^\pi$ at $2^-$ \cite{Doornenbal}, although the authors further encourage its exact determination.

Using the total spin-parity to be $2^-$ for the ground state of $^{34}$Na [$^{33}$Na$(3/2^{+})$ $\otimes$ \textit{$2p_{3/2}\nu$}] as predicted by Ref. \cite{Doornenbal} (unless specified otherwise), we present here the results when a $^{34}$Na projectile, presumed with an incident beam energy of 100 MeV/u, breaks up elastically off a $^{208}$Pb target to give off $^{33}$Na and a valence neutron as substructures for a three-body problem in the final channel. The one neutron separation energy, $S_n$, for studies when it was not treated as a varied parameter, was fixed at 0.17 MeV \cite{Gaudefroy}. The beam energy was assumed to be 100 MeV/u to ensure forward angle domination of ejected projectile fragments and at the same time, negate any higher order effects like post acceleration \cite{BBHST}. At higher beam energies, the detection of particles becomes easier and forward angle prevalence ensures that the reaction is Coulomb dominated and the pure nuclear contribution as well as its interference effects contributing to the breakup cross-sections are negligible \cite{RC2007PRC}.

For our theory to work, we require that the reaction be dominated only by a single multipolarity. To check whether indeed that is the case, we calculated the total Coulomb dissociation cross-section ($\sigma_{-1n}$) for two multipolarities - \textit{E1} and \textit{E2} - using the Alder-Winther theory \cite{AW}. It was found that the \textit{E1} contribution to the total Coulomb dissociation cross-section was 1.743 barns, whereas the \textit{E2} contribution to the same was only 0.167 millibarns. But constructing the continuum states to study multipole responses is a difficult task in perturbative theories (cf. Fig. 15 of Ref. \cite{TB}). This is not a problem with our post form theory as it includes the target-fragment electromagnetic interaction to all orders as well as the entire non-resonant continuum for all multipoles. Nevertheless, using the Alder-Winther theory, we have checked that indeed, we have the dominance of a single multipolarity and thus, we can use the relative energy spectra results from the FRDWBA theory to calculate the capture cross-sections and eventually, the reaction rates, as discussed in the formalism.

%We use the relative energy spectra results from FRDWBA theory to find the photodisintegration cross-section, which was then utilised to determine the capture cross-sections and eventually, the reaction rates.
% and at the same time, negate any relativistic effects that might be produced \cite{DAE2013}

\subsection{The capture cross-section}
\label{sec:3A}

%Fig. \ref{fig: cap_theta} presents the capture cross-section for the $^{33}$Na(n,$\gamma)^{34}$Na reaction while we vary the projectile-target scattering angle. The calculation were done till the scattering angle reached the grazing angle ($\theta_{at} = 2.27$\textdegree) while we kept the one neutron separation energy to 0.17 MeV and the quadrupole deformation to 0.0. As expected, the cross-section is higher for larger projectile-target scattering angle.

\begin{figure}[ht]
\centering
\includegraphics[height=7.03cm, clip,width=0.465\textwidth]{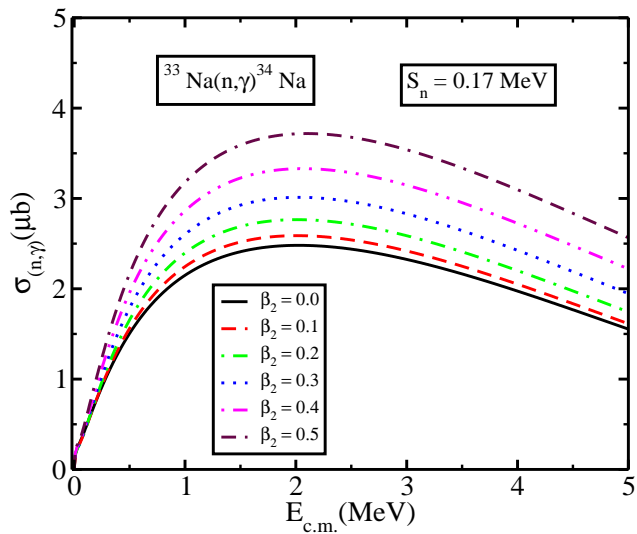}
\caption{\label{fig: cap_beta} (Color online) The capture cross-section for $^{33}$Na(n,$\gamma)^{34}$Na reaction for different values of deformation parameter, $\beta_2$, with the valence neutron separation energy, S$_n$ = 0.17 MeV.}
\end{figure}

\begin{figure}[ht]
\centering
\includegraphics[height=7.03cm, clip,width=0.465\textwidth]{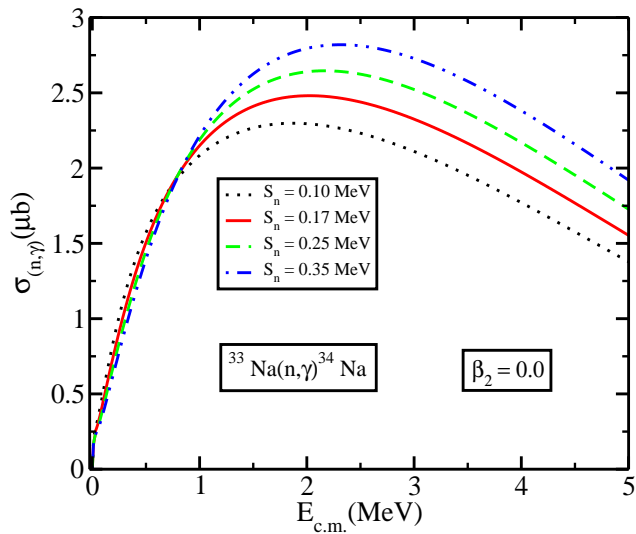}
\caption{\label{fig: cap_Sn} (Color online) The capture cross-section for $^{33}$Na(n,$\gamma)^{34}$Na reaction for various values of valence neutron binding energy with deformation parameter, $\beta_2$ = 0.0. The dotted line is for $S_n$ = 0.10 MeV while the solid, dashed and dash double-dotted lines are for $S_n$ = 0.17, 0.25 and 0.35 MeV, respectively. It is clearly seen that the cross-section values are larger for lower binding energies, a trend which is seen to reverse itself comprehensibly after a c.m. energy $\sim$ 0.75 MeV.}
\end{figure}

In Fig. \ref{fig: cap_beta}, we show the total capture cross-section as obtained in the radiative capture of a neutron by $^{33}$Na. The curves were obtained using the relative energy spectra in conjunction with Eq. (\ref{a1.5}) of the section above. The solid line corresponds to the case of a spherical $^{34}$Na. The cross-section tends to increase with increase in the deformation \footnote{Of course, one needs to remember in hindsight that for the ground state of $^{34}$Na having the configuration $^{33}$Na$(3/2^{+})$ $\otimes$ \textit{$2p_{3/2}\nu$}, we have assumed a spectroscopic factor of 1 \cite{34Na}.}.

Fig. \ref{fig: cap_Sn} shows the same capture cross-section with now the valence neutron binding energy, $S_n$, as a parameter. The $^{34}$Na nucleus was assumed to have a spherical shape for these calculations. The dotted and the solid lines show the results for $S_n$ values 0.10 MeV and 0.17 MeV, whereas the dashed and dash double-dotted lines represent those for $S_n$ values of 0.25 MeV and 0.35 MeV, respectively. Evidently, when $E_{c.m.}$ is above 1 MeV, the $^{34}$Na nucleus profiles with higher binding energies have comparatively higher cross-sections. This is indeed what one would anticipate: that a capture to a state of higher binding energy is more probable. However, when $E_{c.m.}$ goes below 1 MeV, we observe a reversal and the cross-section goes slightly higher for lower values of $S_n$ for a c.m. energy of and lower than $\sim$ 0.75 MeV. Moreover, the difference in cross-section values, although very small, is still not negligible. This is an interesting phenomenon as it falls in the range of the c.m. energy that is responsible for most of the contribution to the reaction rates (as will be seen later), which could ultimately affect the abundance of the nucleus in question. The flip in the capture cross-section with changing neutron separation energy is also vital because the \textit{r-}process paths are actively dependent on the $S_n$ values favoured by the neutrino driven winds \cite{Tera}. In what follows, we shall try to explain the cause of this inversion.

\begin{figure}[ht]
\centering
\includegraphics[height=7cm, clip,width=0.435\textwidth]{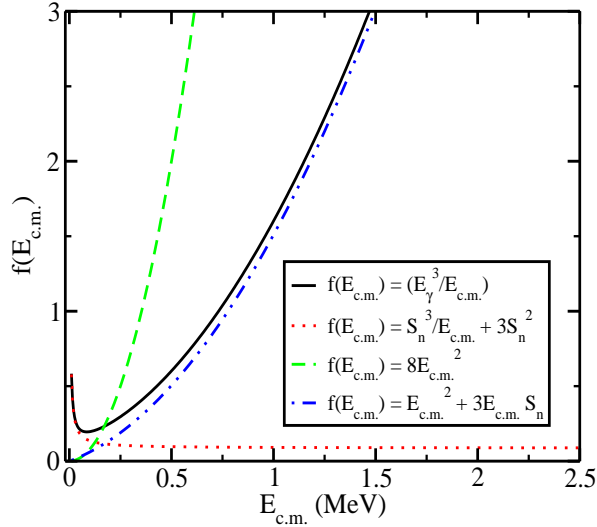}
\caption{\label{fig: f_E} (Color online) A plot of the kinematic factor under different limits of the c.m. energy with respect to the one neutron binding energy. The solid line represents the actual kinematic factor, whereas the dotted, dashed and dash double-dotted curves show the $E_{c.m.} << S_n, E_{c.m.} \simeq S_n$, and $E_{c.m.} >> S_n$ cases, respectively. For details, see text.}
\end{figure}

As explained in Section \ref{sec:2} above, Eqs. (\ref{a1.4}) and (\ref{a1.5}) relate the relative energy spectrum of a two-body breakup reaction with the photodisintegration cross-section, which is then used to obtain the neutron capture cross-section. Combining Eq. (\ref{a1.4}) with Eq. (\ref{a1.5}), we obtain:

\begin{eqnarray}
\sigma_{(n, \gamma)} &=& \left[\dfrac{2{\hat{j_a}}^2}{{\hat{j_b}}^2{\hat{j_b}}^2} \left(\dfrac{1}{2\mu_{bc}}\right)\right] \left[\dfrac{(E_{c.m.} + S_n)^3}{E_{c.m.}}
\left(\dfrac{d\sigma}{dE_{c.m.}}\right)\dfrac{1}{n_{E1}}\right] 	\label{a2.4}
\end{eqnarray}

with $\mu_{bc}$ being the reduced mass of the \textit{b-c} system, which, when expressed in terms of energy units, absorbs the factor of speed of light.

Extracting the kinematic factor from Eq. (\ref{a2.4}),
i.e., $f(E_{c.m.}) = E_{\gamma}^3/E_{c.m.} = \left[\frac{S_n^3}{E_{c.m.}} + 3S_n^2 + 3E_{c.m.}S_n + E_{c.m.}^2 \right] $, we study its behaviour in Fig. \ref{fig: f_E} (the solid curve) for three limiting cases:

(\textit{1}) When $E_{c.m.} << S_n$. Then, we have $f(E_{c.m.}) = \left[\frac{S_n^3}{E_{c.m.}} + 3S_n^2 + \mathcal{O}(E_{c.m.})\right]$. This is depicted by the dotted line in Fig. \ref{fig: f_E}.

(\textit{2}) When $E_{c.m.} \simeq S_n$. In this case, we have $f(E_{c.m.}) = 8E_{c.m.}^2$, which is shown by the dashed curve in the figure.

(\textit{3}) When $E_{c.m.} >> S_n$. We have $f(E_{c.m.}) = \left[ E_{c.m.}^2 + 3E_{c.m.}S_n +\mathcal{O}(S_n^2)\right] $ shown by the dash double-dotted line.

One can clearly see that the actual curve of the kinematic factor changes according to the limiting conditions and this is important in explaining the trend reversal of Fig. \ref{fig: cap_Sn}. The dashed line crosses the actual curve at exactly 0.17 MeV - the value taken for the one neutron binding energy for our calculations. As this one neutron separation energy is indeed very low, the actual trend closely begins to follow condition (\textit{3}) even at a small c.m. energy of 0.5 MeV.

% %\centering
%\includegraphics[height=8.3cm, clip,width=0.65\textwidth]{fe_diff_sn}
%\includegraphics[width=11.0cm]{Jacobi_coordinates}
%\caption{\label{fig: 7} (Colour online) Variation of the kinematic factor in the calculation of the capture cross-section for different values of valence neutron binding energy. The $^{34}$Na nucleus was assumed to be spherical in shape.}
%\end{figure}

\begin{figure}[ht]
\centering
\includegraphics[height=8.3cm, clip,width=0.465\textwidth]{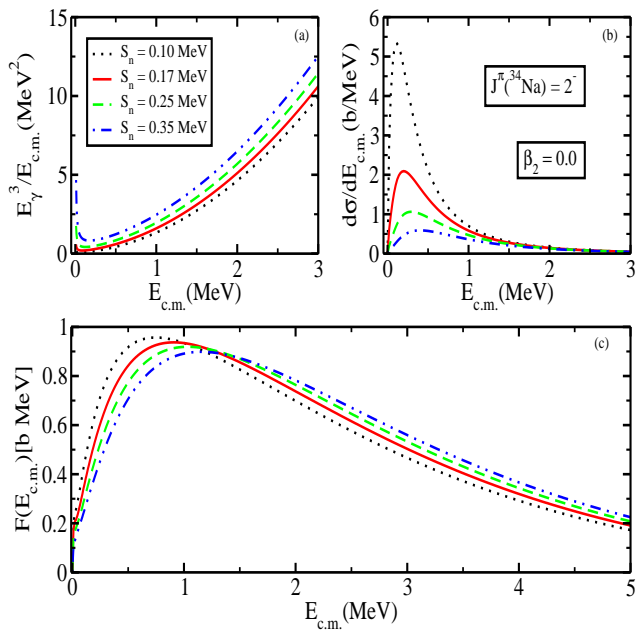}
\caption{\label{fig: FE} (Color online) (a) Variation of the kinematic factor in the calculation of the capture cross-section from photodisintegration cross-section for different values of valence neutron binding energy. The $^{34}$Na nucleus was assumed to be spherical in shape. (b) The relative energy spectra of $^{34}$Na breaking elastically on $^{208}$Pb at 100 MeV/u beam energy due to Coulomb dissociation. (c) Product of the curves in (a) and (b) as per Eq. (\ref{a2.4}). The product gives the reduced capture cross-section which is seen to be higher for lower binding energies up to an $E_{c.m.}$ $\sim$ 1.1 MeV, after which the trend reverses.}
\end{figure}

Since the transposition of the trend in Fig. \ref{fig: cap_Sn} occurs for different values of one neutron separation energy, it is sensible to plot the kinematic factor in Eq. (\ref{a2.4}) for different $S_n$ values, which we show in Fig. \ref{fig: FE}(a). The kinematic factor is seen to increase with increase in the $S_n$ value. Also, it follows more and more the pattern of limiting case (\textit{1}) for Fig. \ref{fig: f_E} above, which in any case is expected as the binding energy becomes larger. Another term crucial in the right hand side of Eq. (\ref{a2.4}) is that of the relative energy spectrum ($d\sigma/dE_{c.m.}$). Fig. \ref{fig: FE}(b) presents the relative energy spectra for $^{34}$Na presumably impinging on $^{208}$Pb at 100 MeV/u beam energy and undergoing elastic breakup due to Coulomb effects. The deformation parameter, $\beta_2$, was set to 0.0. One can notice that the crest of the relative energy spectrum decreases in height with increase in the $S_n$ value of the projectile. The shifting of the peak position of the spectrum towards higher centre of mass energy is also noticeable. It is appropriate to mention here that the peak positions of the relative energy spectra are important as they can be used with scaling properties to get a heuristic estimate of the binding energy of a loosely bound nucleus such as $^{34}$Na \cite{34Na, RCEPJ}.

%\begin{figure}[h]
%\centering
%\includegraphics[height=8.3cm, clip,width=0.65\textwidth]{fe_relen_A}
%\includegraphics[width=11.0cm]{Jacobi_coordinates}
%\caption{\label{fig: 9} (Colour online) The convolution of the kinematic factors with the relative energy spectra to explain the shape of the reduced capture cross-section for the $^{33}$Na(n,$\gamma)^{34}$Na reaction. The shape of $^{34}$Na was taken to be spherical while the valence neutron binding energy was fixed at 0.17 MeV.}
%\end{figure}

Nevertheless, the kinematic factor keeps on increasing monotonically with the c.m. energy of the projectile fragments (as it begins to follow condition (\textit{3}) mentioned above) while the relative energy spectrum initially rises steeply and then has a gradual negative slope. Meanwhile, what matters in Eq. (\ref{a2.4}) is the product of the two functions:\\ $ F(E_{c.m.}) = \left( \dfrac{E_{\gamma}^3}{E_{c.m.}}\right) \times\left( \dfrac{d\sigma}{dE_{c.m.}}\right) $.

Fixing the binding energy at $S_n$ = 0.17 MeV, in Fig. \ref{fig: FE}(c) we show the convolution of the kinematic factor with the relative energy spectrum, which results in a curve that at first increases due to the peak of the relative energy spectrum at lower c.m. energies. However, as the c.m. energy increases, the relative energy spectrum is now negligible (for $E_{c.m.} >$ 2 MeV), and although the kinematic factor increases sharply, the product starts to decrease again. This gives us the preliminary shape of the capture cross-section curves or the \textit{reduced capture cross-section}.

Moreover, what is critical here is that the flip in the trend of the cross-section at $E_{c.m.} \sim$ 1.1 MeV is now clearly evident with increase in the binding energy values. This happens because at the lower c.m. energies, the pattern is dominated by the $(d\sigma/dE_{c.m.})$ curve, and since the difference in the amplitudes of the relative energy spectra is significantly larger (with higher binding energies having smaller amplitudes), the reduced capture cross-section is lower for higher binding energies. As the c.m. energy increases beyond $\sim$ 1.1 MeV, the relative energy spectra begins to merge (as is perceptible from Fig. \ref{fig: FE}(b)) whereas the kinematic factor experiences no such union. Hence, the effect of the kinematic factor begins to dominate, causing a flip in the trend and now lower $S_n$ values have a lower reduced capture cross-section.

\begin{figure}[ht]
\centering
\includegraphics[height=8.3cm, clip,width=0.486\textwidth]{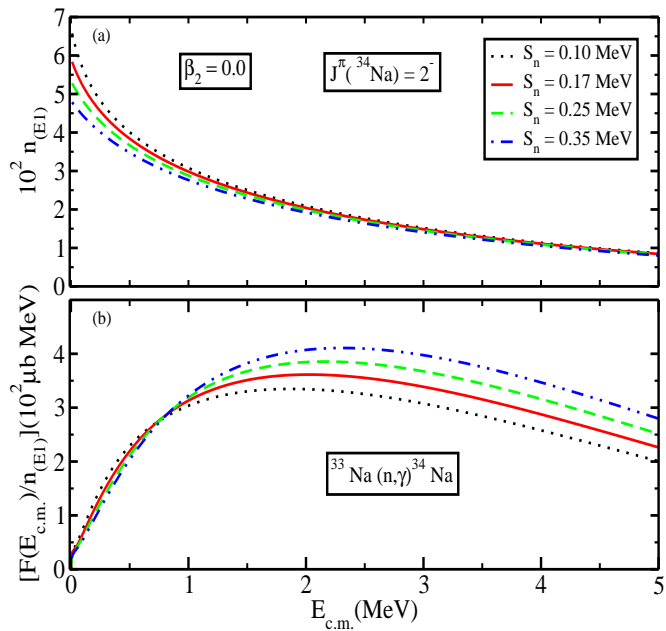}
\caption{\label{fig: nE_sig} (Color online) (a) The virtual photon number as a function of the centre of mass (c.m.) energy of the two photo-dissociated fragments, for different one neutron separation energies with a deformation parameter, $\beta_2$, set to 0.0. (b) Capture cross-section as obtained when the reduced cross-section of Fig. \ref{fig: FE}(c) is divided by with the virtual photon number. Except for the constant coefficients affecting the amplitude (cf. Eq. (\ref{a2.4})), the figure is identical to Fig. \ref{fig: cap_Sn}.}
\end{figure}

In Fig. \ref{fig: nE_sig}(a), we show the virtual photon numbers ($n_{E1}$) against the c.m. relative energy of the core and valence nucleon for various binding energies for the above mentioned breakup reaction. Although the numbers for the virtual photons seem to converge at the higher end of c.m. energy, for relative energies $<$ 2 MeV, higher one neutron binding energies tend to have a significantly lower number of virtual photons.

Fig. \ref{fig: nE_sig}(b) displays the capture cross-section without the constants of Eq. (\ref{a2.4}) (the reduced mass, $\mu_{bc}$ and the spin factors). Having the dimensions of the reduced capture cross-section, it was obtained by dividing the reduced capture cross-section of Fig. \ref{fig: FE}(c) with the virtual photon number shown in Fig. \ref{fig: nE_sig}(a). Because it appears in the denominator in Eq. (\ref{a2.4}), the significant variation in the photon number at lower c.m. energies causes the variation in the total capture cross-section to decrease significantly. It is aptly transparent that Fig. \ref{fig: nE_sig} is identical to Fig. \ref{fig: cap_Sn} apart from the constants of multiplication.

It is noteworthy that since the flip in the calculations of the capture cross-section occurs in the c.m. energy range corresponding to the astrophysically relevant temperature domain, it becomes significant in the behaviour of the reaction rates as will be seen below.

\subsection{Reaction rates}
\label{sec:3B}

\begin{figure}[ht]
\centering
\includegraphics[height=7.03cm, clip,width=0.465\textwidth]{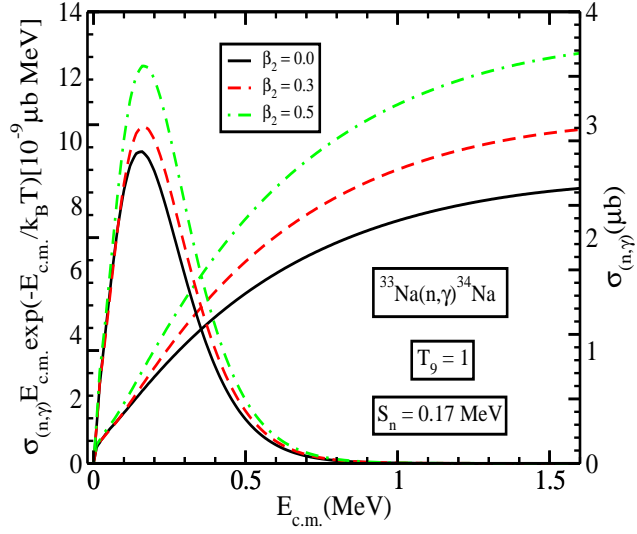}
\caption{\label{fig: cap_int} (Color online) Comparison of capture cross-section for $^{33}$Na(n,$\gamma)^{34}$Na reaction with the integrand involved in the rate of the reaction (Eq. \ref{a1.7}). The right panel of the \textit{y}-axis from 0-4 units gives the capture cross-section which is seen to increase with the c.m. energy. The left panel with values from 0-14 units represents the integrand which is seen to be negligible after a c.m. energy of 1 MeV. The solid lines correspond to a deformation parameter, $\beta_2$ = 0.0, while the dashed and the dash-dotted lines are for $\beta_2$ = 0.3 and 0.5, respectively.  For details, see text.}
\end{figure}

Having studied the capture cross-section and its variations with the one neutron binding energy and the quadrupole deformation, we now proceed to the reaction rates. From Eq. (\ref{a1.7}), it is suitably clear that for a given temperature of the stellar plasma, the rate of a reaction is mainly dependent on the integrand involving the reaction cross-section and the relative energy. This gets support if one checks the contribution of this quantity and plots it with the c.m. energy, something we do in Fig. \ref{fig: cap_int}. The temperature was fixed at $T_9 = 1$. The figure shows that the integrand is substantial only for very small values of c.m. energies (roughly from 0.05 to 0.75 MeV). At such low relative energies, it is seriously difficult to carry out experiments to measure radiative reaction cross-section by direct measurements and that is why one has to resort to indirect methods like CD to calculate the reaction rates. The figure also shows that although the capture reaction cross-section increases for higher c.m. energy values, the confinement of the integrand within the low energy range gives us an idea about the scope of the cross-section contributing chiefly towards the reaction rates. This substantiates why the flip in the cross-section at the lower c.m. energy domain is so important: because it can, in principle, affect the rates in a manner not intuitively thought of. The integrand variation for different deformations of the $^{34}$Na nucleus (viz., $\beta_2$ = 0.0, 0.3, and 0.5, which are represented by the solid, dashed, dash-dotted lines, respectively), is also exhibited. One can see that higher the deformation, higher is the contribution of the integrand to the reaction rate. The one neutron separation energy was once again fixed at 0.17 MeV for these calculations.

%\begin{figure}[h]
%\centering
%\includegraphics[height=8.3cm, clip,width=0.65\textwidth]{int_diff_beta}
%\caption{\label{fig: int_beta} (Colour online) The integrand of the reaction rate expression as a function of c.m. energy for the $^{33}$Na(n,$\gamma)^{34}$Na reaction at different values of deformation parameter, $\beta_2$, for a fixed value of one neutron separation energy, S$_n$ = 0.17 MeV. The temperature in T$_9$ units (10$^{9}$K) was taken to be 1.}
%\end{figure}

\begin{figure}[ht]
\centering
\includegraphics[height=8.3cm, clip,width=0.484\textwidth]{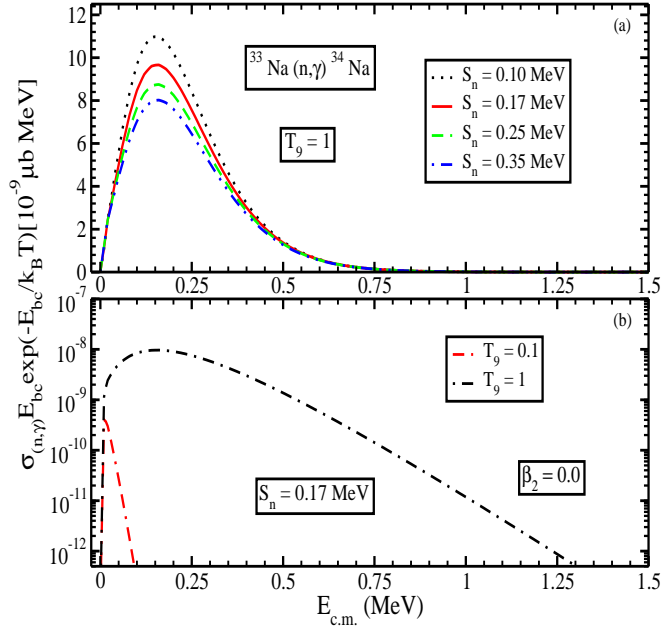}
\caption{\label{fig: int_sn_t9} (Color online) (a) The integrand of the reaction rate expression (Eq. \ref{a1.7}) as a function of c.m. energy for the $^{33}$Na(n,$\gamma)^{34}$Na reaction at different values of one neutron separation energy, S$_n$, for a fixed value of deformation parameter, $\beta_2$ = 0.0. The temperature in T$_9$ units (10$^{9}$K) was taken to be 1. (b) The same at different values of temperature in T$_9$ units for fixed values of deformation parameter, $\beta_2$ = 0.0, and one neutron separation energy, S$_n$ = 0.17 MeV. The double-dash dotted line is for T$_9$ = 0.1 and the dash-dotted line for T$_9$ = 1.}
\end{figure}

%\begin{figure}[h]
%\centering
%\includegraphics[height=8.3cm, clip,width=0.65\textwidth]{int_diff_Sn}
%\includegraphics[width=11.0cm]{Jacobi_coordinates}
%\caption{\label{fig: int_sn} (Colour online) The integrand of the reaction rate expression as a function of c.m. energy for the $^{33}$Na(n,$\gamma)^{34}$Na reaction at different values of one neutron separation energy, S$_n$, for a fixed value of deformation parameter, $\beta_2$ = 0.0. The temperature in T$_9$ units was taken to be 1.}
%\end{figure}

Fig. \ref{fig: int_sn_t9}(a) shows the behaviour of the integrand when the one neutron binding energy, $S_n$, was varied while the deformation parameter was kept fixed at 0.0 and the stellar temperature, in T$_9$ units, was taken to be 1. That higher separation energy tends to lower the peak value of the integrand is the inference from this curve. This should not be surprising since the integrand is mainly dependent on the capture cross-section, which follows a similar pattern for lower c.m. energies as shown above in Fig \ref{fig: cap_Sn}.

%\begin{figure}[ht]
%\centering
%\includegraphics[height=7.03cm, clip,width=0.45\textwidth]{Int_diff_t9}
%\includegraphics[width=11.0cm]{Jacobi_coordinates}
%\caption{\label{fig: int_t9} (Color online) The integrand of the reaction rate expression as a function of c.m. energy for the $^{33}$Na(n,$\gamma)^{34}$Na reaction at different values of temperature in T$_9$ units (10$^{9}$K) for fixed values of deformation parameter, $\beta_2$ = 0.0, and one neutron separation energy, S$_n$ = 0.17 MeV. The solid line is for T$_9$ = 0.1 and the dash-double dotted line for T$_9$ = 1.}
%\end{figure}

Since we have studied the behavior of the integrand with variation in one neutron separation energy and quadrupole deformation, it would not be unwise to study it with variation in temperature. This is precisely what is shown in Fig. \ref{fig: int_sn_t9}(b). As the equilibrium temperature in our case study is $T_9$ = 0.62, we restrict ourselves to examine the integrand response only at $T_9$ equal to 0.1 and 1. We find that at the lower limit of $T_9 = 0.1$, the integrand peak (double-dash dotted line) is orders of magnitude smaller in comparison to the curve for $T_9 = 1$ (dash-dotted line). In fact, it peaks only for an extremely small c.m. energy range. However, this huge variation is to be expected as the integrand depends on the temperature exponentially. %It is worth noting here that a $T_9$ in the range of 0.05 \textendash 10 refers to an $E_{c.m.}$ approximately in the range of 50 keV to 1 MeV which supplements our decision to check the behaviour of the integrand for a fixed $T_9$ (= 1).

%\begin{figure}[ht]
%\centering
%\includegraphics[height=8.3cm, %clip,width=0.65\textwidth]{rate_diff_beta2}
%\includegraphics[width=11.0cm]{Jacobi_coordinates}
%\caption{\label{fig: rate_beta} (Colour online) Capture reaction rates for $^{33}$Na(n,$\gamma)^{34}$Na reaction for different values of deformation parameter. The one neutron separation energy was fixed at 0.17 MeV.}
%\end{figure}

%\begin{figure}[ht]
%\centering
%\includegraphics[height=8.3cm, clip,width=0.65\textwidth]{rate_Sn_3}
%\includegraphics[width=11.0cm]{Jacobi_coordinates}
%\caption{\label{fig: rate_sn} (Colour online) The capture reaction rates for $^{33}$Na(n,$\gamma)^{34}$Na reaction as a function of temperature in units of 10$^{9}$K (T$_{9}$) for different values of S$_{n}$ for a spherical nucleus.}
%\end{figure}

We now discuss the reaction rates for neutron capture and $\alpha$-capture reactions by $^{33}$Na. A comparison of the $\alpha$-capture rates with that of the neutron capture is crucial in determining if the ($\alpha$,n) reaction will dominate over (n,$\gamma$) and halt the \textit{r-}process path flow towards the neutron drip line, thereby favoring the production of matter with a higher proton number. It was predicted in Ref. \cite{Tera} that sodium isotopes maintain a strong flow towards the drip line by neutron capture reactions. This flow, nonetheless, could be broken if the competing $\alpha$-capture rate is more than the neutron capture.

\begin{figure}[ht]
\centering
\includegraphics[height=12cm, clip,width=0.484\textwidth]{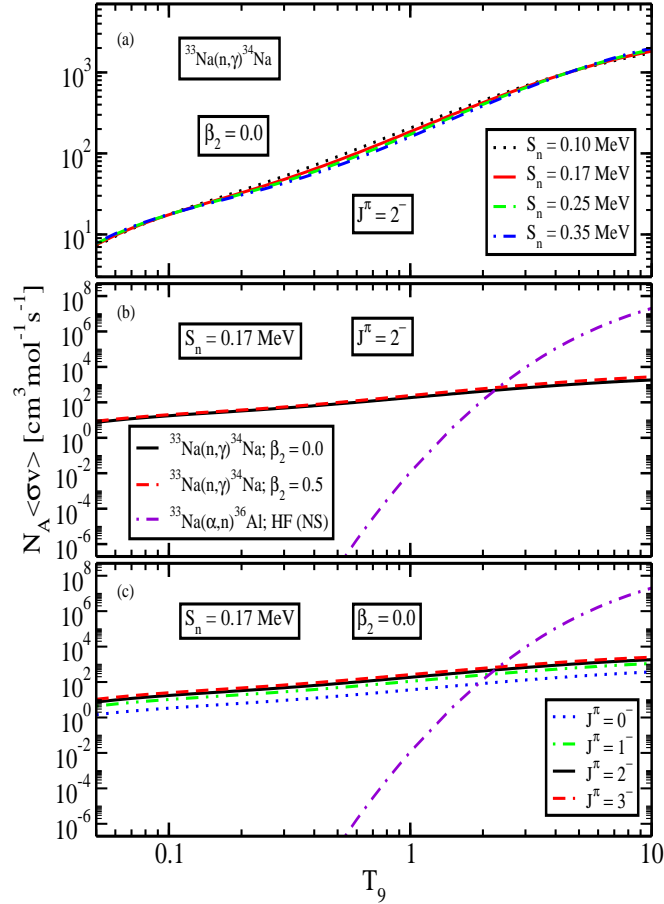}
\caption{\label{fig: rates} (Color online) (a) The capture reaction rates for $^{33}$Na(n,$\gamma)^{34}$Na reaction as a function of temperature in units of 10$^{9}$K (T$_{9}$) for different values of $S_{n}$ for a spherical nucleus having g.s. spin-parity as $2^-$. The legend scheme is same as that of Fig. \ref{fig: cap_Sn}. (b) Capture reaction rates for $^{33}$Na(n,$\gamma)^{34}$Na reaction using CD method with deformation parameter values 0.0 (solid line) and 0.5 (dashed line), and for $^{33}$Na($\alpha$,n)$^{36}$Al reaction (dash-dotted line) using HF theory calculated from the NON-SMOKER code \cite{NS}. The one neutron separation energy was fixed at 0.17 MeV and g.s. $J^\pi$ taken to be $2^-$. (c) Rates of the same reactions as in (b) above with now the $S_n$ and $\beta_2$ fixed for various g.s. spin-parities of $^{34}$Na ($0^-, 1^-, 2^-$ and $3^-$, shown by dotted, dash double-dotted, solid and dashed lines, respectively). From (b) and (c), it is evident that at the equilibrium temperature, T$_9$ = 0.62, the rate for the neutron capture is far more than the rate for the alpha capture by $^{33}$Na.}
\end{figure}

Shown in Fig. \ref{fig: rates}(a) are the reaction rates per mole obtained from the reaction cross-sections as given by Eq. (\ref{a1.7}) for the $^{33}$Na(n,$\gamma)^{34}$Na reaction for a spherical $^{34}$Na nucleus and various values of its one neutron binding energy. The rate varies from about 15 cm$^3$mol$^{-1}$s$^{-1}$ to about 1500 cm$^3$mol$^{-1}$s$^{-1}$ as $T_9$ rises from 0.1 to 10. The value around the equilibrium temperature, $T_9 = 0.62$, is about 80 cm$^3$mol$^{-1}$s$^{-1}$ at an $S_n$ value of 0.17 MeV. The lower binding energy configurations of $^{34}$Na appear to have a slightly higher reaction rate in complete agreement with the trend observed in the capture cross-section. However, the slight, although non-negligible, variation of the reaction rate with small changes in the one neutron separation energy once again points out to the vitality of knowing this energy with precision and accuracy.

Fig. \ref{fig: rates}(b) shows a comparison of the reaction rates for the cases when the $^{33}$Na nucleus captures a neutron and an $\alpha$ particle, i.e., for the $^{33}$Na(n,$\gamma)^{34}$Na and $^{33}$Na($\alpha$,n)$^{36}$Al reactions for the same astrophysically relevant stellar temperature range ($T_9 = 0.05-10$). For the (n,$\gamma$) rates, the neutron separation energy was fixed at 0.17 MeV and the outputs for calculations done for deformation values of 0.0 and 0.5 are plotted. The rate for the ($\alpha$,n) reaction was obtained from the Hauser-Feshbach theory using the NON-SMOKER code. HF is a widely used statistical theory to calculate capture rates for astrophysical purposes, though it may not be very precise for exotic nuclei due to the uncertainties involved in the model \cite{Bertolli,NS}. Nevertheless, apart from their easy availability, these estimates obtained from it can be used because of the uncertainties being smaller than the difference of the rates between ($\alpha$,n) and (n,$\gamma$) reactions. 

As is evident, for $T_9 \leq 1$, although there is hardly any significant difference between the rates of neutron capture by a spherical and a deformed $^{34}$Na nucleus ($\beta_2$ = 0.0 and 0.5, respectively), the neutron capture reaction dominates over the $\alpha$-capture. In fact, at the equilibrium temperature of $T_9 = 0.62$, the neutron capture rate outscores the $\alpha$-capture by more than six orders of magnitude. Thus, in this temperature region, the classical \textit{r}-process path flow involving $\beta$-decay after the (n,$\gamma$)-($\gamma$,n) reactions, has more probability. However, as the temperature increases, the neutron capture does not pick up speed as much as the $\alpha$-capture and for $T_9 > 2$, the rate for the $\alpha$-capture is more and dominating, pointing to the reasoning that above these temperatures, the elements with a higher atomic number are more probable to form via the $\alpha$ induced processes. Fig. \ref{fig: rates} also corroborates that the reliance of the reaction rate on both the $\beta_2$ and the $S_n$ follows the trends observed in the dependence of the capture cross-section on these parameters.

In Fig. \ref{fig: rates}(c), a similar comparison of the two capture reaction rates is made when the ground state spin of $^{34}$Na is varied for constant $S_n$ (= 0.17 MeV) and $\beta_{2}$ (= 0.0) values. The dotted, dash double-dotted, solid and dashed curves refer to the g.s. spin of $0^-, 1^-, 2^-$ and $3^-$, respectively. The calculations show that higher the spin, higher is the rate. This is due to the spin coefficient factor entering Eq. (\ref{a1.5}). Although the difference in the rates is fairly appreciable, it is still not substantial in comparison to the difference in the rate of neutron capture and the $\alpha$-capture by $^{33}$Na, the latter being displayed by the dash-dotted line as in Fig. \ref{fig: rates}(b).

Thus, the predictions made by Ref. \cite{Tera} seem to hold their ground in case of the \textit{r}-process path flow being towards the drip line for Na isotopes. But there is a need to further verify these results from experimental observations and CD can be an important tool in that quest as CD experimental results can be used to that effect.

%%%%%%%%%%%%%%%%%%%%%%%%%%%%%%%%%%%%%%%%%%%%%%%%%%%%%%%%%%%%%%%%%%%%%%%%%%%%%%%%%%%%%%%%%%%%%%%%%%%%%%%%
\section{Conclusions}
\label{sec:4}

In the route towards the creation of seed nuclei for the \textit{r}-process, neutron capture reactions in the medium mass region (\textit{N} = 20-30) could push elemental abundances towards the neutron drip line by being more prominent than their $\alpha$-capture counterparts. We have investigated the role of $^{34}$Na in this process to see if it really does follow this pattern. % as was done in the case of $^{37}$Mg \cite{36Mg}.

%With a short dynamic timescale for the \textit{r}-process, the neutrons are not freely available to form heavier nuclei as they are consumed in neutron capture reactions to push nucleosynthesis towards the neutron drip line from the valley of stability. Hence, abundances of neutron rich light and medium mass isotopes can, in part, be ascribed to the neutron capture reactions being more prominent than their $\alpha$-capture counterparts, leading to these exotic nuclei playing pertinent characters as seeds for \textit{r}-process nucleogenesis. 

We have used the method of Coulomb dissociation through our theory of finite range distorted wave Born approximation amplified to incorporate the consequences of deformation, and studied the theoretical elastic Coulomb breakup of $^{34}$Na on $^{208}$Pb at 100 MeV/u beam energy to give off a $^{33}$Na core and a valence neutron. Around this beam energy ($\sim$ a few hundred MeV/u), the final channel fragments emanate with higher velocities and are usually easier to detect. With properly chosen measurement conditions, it is possible to study low relative energy outgoing fragments, which can give insights to reactions at the astrophysically important energies of a few keV to a few hundreds of keV range. We have then used the principle of detailed balance to study the reverse capture reaction $^{33}$Na(n,$\gamma)^{34}$Na and calculate its cross-section and reaction rate at the stellar temperature range concerned. This indirect technique had been used in the past to study various capture reactions and their rates \cite{9Li, 16N, 15C}.

We find that an increase in the values of the deformation parameter, $\beta_2$, resulted in a higher capture cross-section. It is noteworthy that this significant change in the capture cross-section occurs even though we have used a spherical wave function and introduced deformation in our theory only via the axially symmetric quadruply deformed potential, $V_{bc}$, appearing in the transition amplitude. Thus, using this approximation is not very unwise, although calculations with a fully deformed wave function would be desirable and welcome. A variation in the one neutron binding energy, $S_n$, showed an interesting response. At low centre of mass energies of the final channel projectile fragments, a higher cross-section for lower binding energy values was obtained. This trend reversed itself at higher centre of mass energies of the fragments but the cause of resulting flip was obtained analytically. It is worth mentioning that knowledge of the exact value of $S_n$ is important not only from the structural point of view, but it is also crucial to understand the \textit{r}-process path flow, as the \textit{r}-process path strongly depends on the $S_n$ value favoured by the neutrino-driven winds.

As the rate integrand manifested, the behaviour of the capture cross-section at the lower centre of mass energy values is central to understanding the conduct of the reaction rates for the relevant astrophysical energy and temperature range. Our calculations suggest that under the specified physical conditions of the stellar plasma (at equilibrium temperature, $T_9 = 0.62$ and mass density, $\rho = 5.4\times10^2$g/cc, where the main path of the \textit{r}-process reaction network goes through extremely neutron rich nuclei), variations in the one neutron separation energy and the deformation parameters do not alter the rate of the $^{33}$Na(n,$\gamma)^{34}$Na reaction drastically, though there is an appreciable change in the rates with changing ground state spin of $^{34}$Na. However, for the competing (n,$\gamma$) and ($\alpha$,n) reactions, the rate for the $^{33}$Na(n,$\gamma)^{34}$Na reaction is highly dominant over the rate for the $^{33}$Na($\alpha$,n)$^{36}$Al reaction. Consequently, the $\alpha$-capture should not break the (n,$\gamma$) \textit{r}-process path for $^{33}$Na isotope. This should effectively push the isotopic abundance of Na isotopes towards the neutron drip line.

Thus, there is a need to determine these reaction rates very accurately for exotic nuclei near the neutron drip line. In fact, between the $^{33}$Na(n,$\gamma)^{34}$Na and $^{33}$Na($\alpha$,n)$^{36}$Al reactions, $^{33}$Na lies at a branching point from where the abundance of the possible seed nuclei could be strongly influenced. Since direct experiments at this energy range are very arduous, for indirect methods, a precise and exact determination of one neutron separation energy of $^{34}$Na along with its g.s. spin-parity should be known to deduce its reaction rates. Ideally, one would desire the experimentally measured dipole response or the relative energy spectra results to have a good understanding on the continuum structure of $^{34}$Na, not much about which is known. We have assumed it to be non-resonant for our calculations, but even for a continuum with narrow resonances, reaction rates can be computed easily, albeit with a different formalism \cite{16N}. Experimental information about the total cross-section for the CD of $^{34}$Na and the momentum distributions of the charged core is also prudently sought to restrict the g.s. properties of this halo nucleus. Therefore, we strongly encourage experiments to put more stringent limits on the uncertain structural parameters of $^{34}$Na (viz., its $J^\pi$, $S_n$ and $\beta_2$) and its relative energy spectrum. Consequently, the resultant capture cross-sections and the rates for the $^{33}$Na(n,$\gamma)^{34}$Na reaction would pave a way to confirm the predictions about its role in the \textit{r}-process reaction network.

\section*{Acknowledgment}
This text results from research supported by the Department of Science and Technology, Govt. of India, (SR/S2/HEP-040/2012). Support from MHRD grant, Govt. of India, to [GS] is gratefully acknowledged. [S] is supported by the U.S. NSF Grant No. PHY-1415656 and the U.S. DOE Grant No. DE-FG02-08ER41533.
%%%%%%%%%%%%%%%%%%%%%%%%%%%%%%%%%%%%%%%%%%%%%%%%%%%%%%%%%

%\begin{acknowledgments}
%This text presents results from research supported by the Department of Science and Technology, Govt. of India, (SR/S2/HEP-040/2012). 
%\end{acknowledgments}


\begin{thebibliography}{99}

\bibitem{BBFH} E. M. Burbidge, G. R. Burbidge, W. A. Fowler, and F. Hoyle, Rev. Mod. Phys. \textbf{29}-4, 547 (1957).

\bibitem{Rolf} C. Rolfs, H. P. Trautvetter, and W. S. Rodney, {Rep. Prog. Phys.} \textbf{50}, 233 (1987).

\bibitem{40BBFH} G. Wallerstein \textit{et al.}, Rev. Mod. Phys. \textbf{69}-4, 995 (1997).

\bibitem{Iliadis} J. Jos\'{e}, and C. Iliadis, {Rep. Prog. Phys.} \textbf{74}, 096901 (2011).

\bibitem{Wiescher} M. Wiescher, J. G\"{o}rres, E. Uberseder, G. Imbriani, and M. Pignatari, Annu. Rev. Nucl. Part. Sci. \textbf{60}, 381 (2010).

\bibitem{Hoffman} R. D. Hoffmann, S. E. Woosely, G. M. Fuller, and B. S. Meyer, Astrophys. J. \textbf{460}, 478 (1996).

\bibitem{Wallace} R. K. Wallace, and S. E. Woosley, Astrophys. J. Suppl. \textbf{45}, 389 (1981).

\bibitem{Cowan} J. J. Cowan, F.-K. Thielemann, and J. W. Truran, Phys. Rep. \textbf{208}, 267 (1991).

\bibitem{Meyer} B. S. Meyer, G. J. Mathews, W. M. Howard, S. E. Woosely, and R. D. Hoffmann, Astrophys. J. \textbf{399}, 656 (1992).

\bibitem{Qian} Y.-Z. Qian, P. Vogel, and G. J. Wasserburg, Astrophys. J. \textbf{506}, 868 (1998).

\bibitem{Thielemann} F.-K. Thielemann \textit{et al.}, Prog. Part. Nucl. Phys. \textbf{66}, 346 (2011).

\bibitem{Tanvir} N. R. Tanvir A. J. Levan, A. S. Fruchter, J. Hjorth, R. A. Hounsell, K. Wiersema, and R. L. Tunicliffe, Nature, \textbf{500}, 547 (2013).

\bibitem{Bauswein} A. Bauswein, R. Ardevol Pulpillo, H.-T. Janka, and S. Goriely, Astrophys. J. \textbf{795}, L9 (2014).

\bibitem{Mennekens} N. Mennekens, and D. Vanbeveren, Astron. Astrophys. \textbf{564}, A134 (2014).

\bibitem{Voort} F. van de Voort, E. Quataert, P. F. Hopkins, D. Kere\v{s}, and C.-A. Faucher-Gigu\`{e}re, Mon. Not. R. Astron. Soc. \textbf{447}, 140 (2015).

\bibitem{Shen} S. Shen, R. J. Cooke, E. Ramirez-Ruiz, P. Madau, L. Mayer, and J. Guedes, Astrophys. J. \textbf{807}, 115 (2015).

\bibitem{Lippuner} J. Lippuner, and L. F. Roberts, Astrophys. J. \textbf{815}, 82 (2015).

\bibitem{Sasaqui} T. Sasaqui, T. Kajino, G. J. Mathews, K. Otsuki, and T. Nakamura, Astrophys. J. \textbf{634}, 1173 (2005).

\bibitem{Tera} M. Terasawa, K. Sumiyoshi, T. Kajino, G. J. Mathews, and I. Tanihata, {Astrophys. J.} \textbf{562}, 470 (2001).

\bibitem{War} E. K. Warburton, J. A. Becker, and B. A. Brown, {Phys. Rev. C} \textbf{41}, 1147 (1990).

\bibitem{Moto} T. Motobayashi \textit{et al.}, Phys. Lett. B \textbf{346}, 9 (1995).

%\bibitem{VT} V. Tripathi, \textit{et al.}, Phys. Rev. Lett. \textbf{101}, 142504 (2008).

\bibitem{Naka2009} T. Nakamura \textit{et al.}, Phys. Rev. Lett. \textbf{103}, 262501 (2009).

\bibitem{31Ne} Shubhchintak, and R. Chatterjee, {Nucl. Phys. A} \textbf{922}, 99 (2014).

\bibitem{37Mg} Shubhchintak, Neelam, R. Chatterjee, R. Shyam, and K. Tsushima, {Nucl. Phys. A} \textbf{939}, 101 (2015).

\bibitem{Doornenbal} P. Doornenbal \textit{et al.} Prog. Theor. Exp. Phys. \textbf{2014}, 053D01 (2014).

\bibitem{Goriely} S. Goriely, Astr. Astrophys. \textbf{325}, 414 (1997).

\bibitem{DataA33} J. Chen, and B. Singh, {Nucl. Data Sheets} \textbf{112}, 1393 (2011).

\bibitem{BaurBert} G. Baur, C. A. Bertulani, and H. Rebel, Nucl. Phys. A \textbf{458}, 188 (1986).

\bibitem{9Li} P. Banerjee, R. Chatterjee, and R. Shyam, {Phys. Rev. C} \textbf{78}, 035804 (2008).

\bibitem{Nakamura} T. Nakamura \textit{et al.}, Phys. Rev. Lett. \textbf{83}, 1112 (1999).

%\bibitem{16NReply} Shubhchintak, Neelam, and R. Chatterjee, {Phys. Rev. C} \textbf{93}, 059802 (2016).

\bibitem{BHTreview} G. Baur, K. Hencken, and D. Trautmann, Prog. Part. Nucl. Phys. \textbf{51}, 487 (2003).

\bibitem{BertJPG} C. A. Bertulani, J. Phys. G: Nucl. Part. Phys. \textbf{25}, 1959 (1999).

\bibitem{Gaudefroy} L. Gaudefroy \textit{et al.}, Phys. Rev. Lett. \textbf{109}, 202503 (2012).

\bibitem{34Na} G. Singh, Shubhchintak, and R. Chatterjee, Phys. Rev. C \textbf{94}, 024606 (2016).

\bibitem{NS} T. Rauscher, At Data Nucl. Data Tables \textbf{79}, 47 (2001): ibid., computer code NON-SMOKER available at: http://nucastro.org/nonsmoker.html.

\bibitem{RC1} R. Chatterjee, P. Banerjee, and R. Shyam, {Nucl. Phys. A} \textbf{675}, 477 (2000).

\bibitem{Hama} I. Hamamoto, {Phys. Rev. C} \textbf{69}, 041306(R) (2004).

\bibitem{VinhMau} N. Vinh Mau,  Nucl. Phys. A \textbf{592}, 33 (1995).

\bibitem{KobayashiPRL} N. Kobayashi \textit{et al.}, Phys. Rev. Lett. \textbf{112}, 242501 (2014).

\bibitem{BertBaur} C. A. Bertulani, and G. Baur, Phys. Rep. \textbf{163}, 299 (1988).

\bibitem{RolfsRodney} C. E. Rolfs, and W. S. Rodney, \textit{Cauldrons in the Cosmos} (University of Chicago Press, Chicago 1988).

\bibitem{16N} Neelam, Shubhchintak, and R. Chatterjee, {Phys. Rev. C} \textbf{92}, 044615 (2015).

\bibitem{Tribble} R. E. Tribble, C. A. Bertulani, M. La Cognata, A. M. Mukhamedzhanov, and C. Spitaleri, Rep. Prog. Phys. \textbf{77}, 106901 (2014).

\bibitem{Casal} J. Casal, M. Rodr\'{\i}guez-Gallardo, J. M. Arias, and J. G\'{o}mez-Camacho, {Phys. Rev. C} \textbf{93}, 041602(R) (2016).

\bibitem{ShubhIOP} C.A. Bertulani, Shubhchintak, A. Mukhamedzhanov, A. S. Kadyrov, A. Kruppa, and D. Y. Pang, J. Phys.: Conf. Ser. \textbf{703}, 012007 (2016).

\bibitem{BBHST} P. Banerjee, G. Baur, K. Hencken, R. Shyam, and D. Trautmann, Phys. Rev. C \textbf{65}, 064602 (2002).

%\bibitem{DAE2013} Gagandeep Singh, Shubhchintak, and Rajdeep Chatterjee, Proceedings of the DAE Symp. on Nucl. Phys. \textbf{58}, 558 (2013).

\bibitem{RC2007PRC} R. Chatterjee, {Phys. Rev. C} \textbf{75}, 064604 (2007).

\bibitem{AW} A. Winther, and K. Alder, Nucl. Phys. A \textbf{319}, 518 (1979).

\bibitem{TB} S. Typel, and G. Baur, {Nucl. Phys. A} \textbf{759}, 247 (2005).

\bibitem{RCEPJ} R. Chatterjee, L. Fortunato, and A. Vitturi, {Eur. Phys. J. A} \textbf{35}, 213 (2008).

%\bibitem{36Mg} Shubhchintak, R. Chatterjee, and R. Shyam, \textit{to be published}.

\bibitem{Bertolli} M. G. Bertolli, T. Kawano, and H. Little, {Nucl. Data Sheets} \textbf{120}, 194 (2014).

\bibitem{15C} Shubhchintak, Neelam, and R. Chatterjee, Pramana-J. Phys. \textbf{83}, 533 (2014).

\end{thebibliography}
%%%%%%%%%%%%%%%%%%%%%%%%%%%%%%%%%%%%%%%%%%%%%%%%%%%%%%%%%%%%%%%%%%%%%%%%%%%%%%%%%%%%%%%%%%%%%%%%%%%%%%%

%%%%%%%%%%%%%%%%%%%%%%%%%%%%%%%%%%%%%%%%%%%%%%%%%%%%%%%%%%%%%%%%%%%%%%%%%%%%%%%%%%%%%%%%%%%%%%%%%%%%%%%%%%%%%%%%%%%%%%%%%%%%%%%%%%%%%%%%%%%%%%%%%%%%%%%%%%%%

\end{document}